\documentclass[a4paper,10pt]{article}

\usepackage{amsmath, amssymb, amsthm, bm}

\usepackage{graphicx}
\usepackage{booktabs}
\usepackage{dcolumn}

\usepackage{xcolor,soul}
\usepackage{paralist}
\usepackage{titlesec}
\usepackage{fancyhdr}
\usepackage{etoolbox}
\usepackage{lineno} 
\usepackage[margin=1in]{geometry}
\usepackage{hyperref}

\title{Actuation of cell layers in three dimensions}
\author{Kirsten Endresen$^{1,\dagger*}$, Aniruddh Murali$^{2,\dagger}$, Birte Geerds$^3$, Daniel J.G. Pearce$^3$, Francesca Serra$^{2,*}$ \\
$^{1}$ Dept. Physics and Astronomy, Johns Hopkins University\\
$^{2}$ Dept. Physics, Chemistry and Pharmacy, University of Southern Denmark\\ 
$^{3}$ Dept. Theoretical Physics, University of Geneva\\
$^{\dagger}$These authors contributed equally to this work. \\
  $^{*}$Corresponding authors: \texttt{kirsten.d.endresen@gmail.com}, \texttt{serra@sdu.dk}
  \date{}
  }

\begin{document}






\maketitle

\begin{abstract}
The alignment of fibers and cells in living tissues affect their mechanical properties and functionality. In this context, one can draw an analogy between tissues and nematic liquid crystal elastomers. We explore this analogy by growing fibroblasts on 2D-patterned substrates and observing the contraction of cell sheets upon detachment from the substrates.
When fibroblast sheets detach, they undergo an anisotropic contraction, with maximum contraction along the nematic director, like nematic elastomers do during phase transition. We quantify this anisotropy using substrates patterned with stripes to induce alignment, finding that cell sheets resemble nematic elastomers with negative Poisson ratio. Then, we apply design principles used for programming curvature in nematic elastomers to actuate 3D structures in the detached fibroblast layers, demonstrating an application of these principles and we support the results with simulations. This proof of concept shows the ability to control the 3D shape through 2D patterning in cell layers, leading to promising avenues to program tissues.
\end{abstract}

\section*{Introduction}

Morphogenesis, the generation of shape in biological tissues, is a complex phenomenon that happens due to the interplay of chemical, biological and physical cues \cite{davies2013morpho, lenne2007morpho}. Evidence is gathering that among the physical cues, cells' nematic alignment plays an important role \cite{Perez2025morpho, ray2024morpho, Balasubramaniam2022, cui2020morpho, guruciaga2024morpho}. In addition, the generation of tissues with desired target shapes through the self-assembly and self-organization of cells is an appealing idea for the creation of implants with maximal adaptability to the target organ.  In tissue engineering the alignment of cells has been highlighted as an important factor~\cite{Gao2020, Li2014}. For example, cardiac implants or other bioprinted tissues improve if the orientation of the cells matches the orientation of the surrounding tissue~\cite{Xu2024, Zhu2017, Kim2012}. The alignment of cells can be controlled in 2D cultures and then used in 3D multi-layered structures~\cite{Okano_tissues, Okano2_engineering_approach, Masuda2016}, an approach that is complementary to others, such as the use of programmable and tunable gels or scaffolds~\cite{Tomba2022,Elvitigala2023}.

The analogy between spontaneously aligning cells and nematic liquid crystals can provide relevant insight~\cite{Saw2018}. Cell alignment influences cell-cell communication, cell migration, and even morphological features and differentiation. For example, by analyzing topological defects in the nematic order of cells, it was shown that they impact the rate of cell death~\cite{Saw2017}, the formation of protrusions from a flat cell layer~\cite{Harding2014, Yara2018, Guillamat2022, Kawaguchi2017, Ho2024, Wang2023} and the potential for tissue regeneration~\cite{Makhija2024}. 

One interesting aspect of this analogy comes into focus when one considers the effect of the liquid crystal order on the mechanical properties of tissues, in particular by considering tissues as a special type of liquid crystal elastomer. Liquid crystal elastomers are polymer networks with liquid crystal mesogens embedded in the polymer main or side chains~\cite{Terentjev2003}. One of the consequences of their mesogenic nature is the ability to change their macroscopic shape in response to a phase transition, due to the coupling between nematic order and mechanical properties.  

This property, often called actuation, is a consequence of the coupling of the mechanical properties with the order parameter of the material. This is exemplified by the simple observation that the equilibrium shape of a uniformly aligned nematic elastomer changes in the passage from the nematic to the isotropic phase. If the elastomer is crosslinked in the aligned nematic phase, at the phase transition it contracts along the nematic director and expands in the directions perpendicular to it. By imposing initially a non-uniform alignment, for example inducing topological defects, this anisotropic expansion/contraction can result in generation of Gaussian curvature~\cite{Aharoni2018, Modes2011, Mostajeran2016}. 

This property is intriguing in the context of cell sheets. In particular, populations of fibroblasts grown in 2D at high density display nematic order~\cite{Duclos2016, Duclos2014}. Moreover, as they proliferate they secrete polymers that create an extracellular matrix, an elastic medium in which the cells are embedded~\cite{DeLeon-Pennell2020}. The nematic order of the cells favors a certain degree of alignment of the elastic matrix~\cite{Li2017}. We could therefore expect that this composite elastic medium behaves like liquid crystal elastomers, where the presence of topological defects is associated with generation of Gaussian curvature. To test this hypothesis, we analyze the behavior of layers of fibroblast cells, whose alignment is controlled by a topographic pattern.

\section*{Results}
\subsection*{Uniaxial Contraction}\label{uniaxial_contraction}

\begin{figure}[t!]
    \centering
    \includegraphics[width=0.80\textwidth]{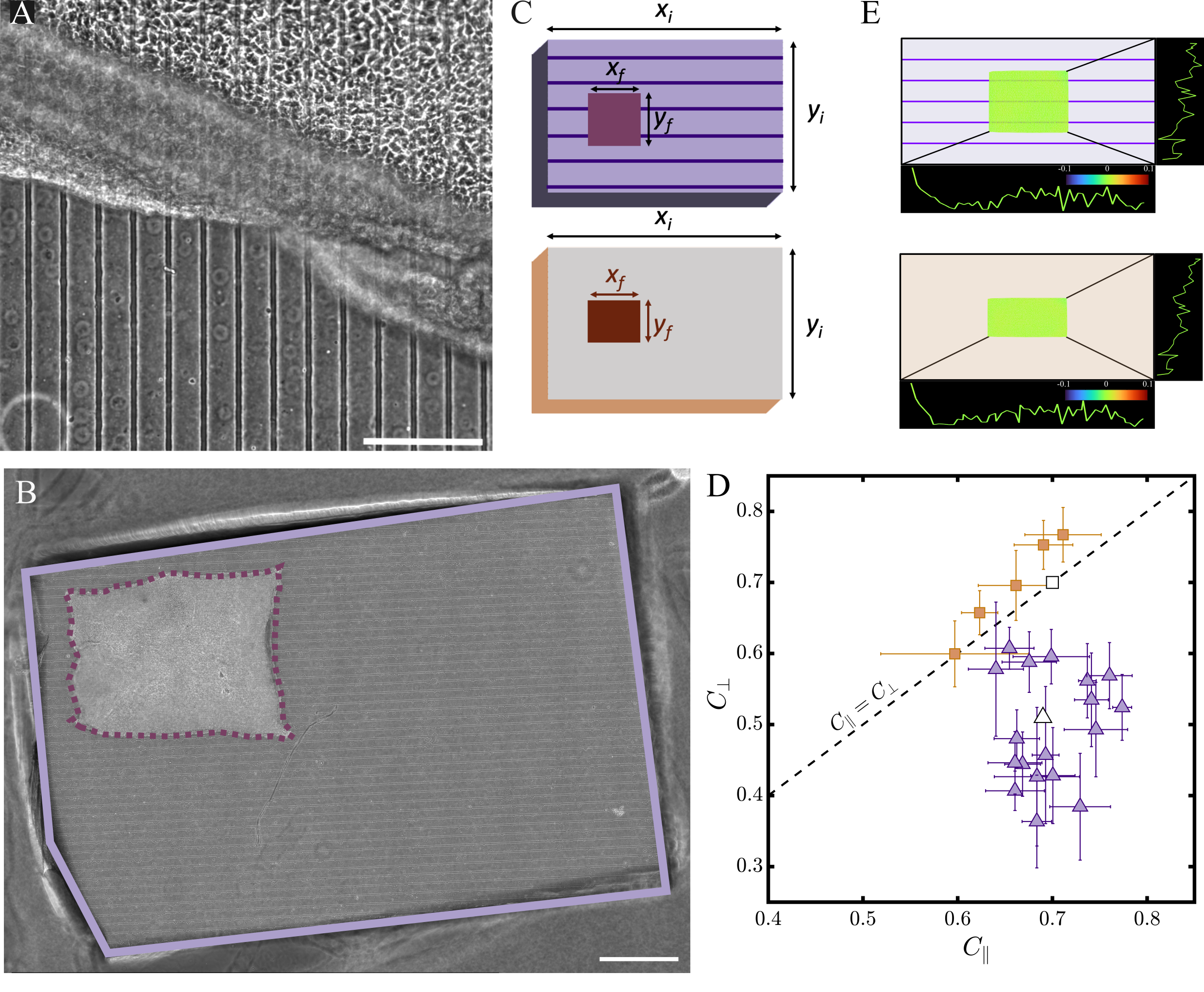}
    \caption{\textbf{Detachment of cell sheets from striped and unpatterned PDMS.} (A) Phase contrast microscopy image of a cell sheet peeling off a PDMS slab with striped pattern (h=2$\mu$m, w=60$\mu$m). The bottom part of the image shows the substrate after cells have peeled off, while in the top part cells are still attached. Scale bar 200$\mu$m. (B) Image of an entire PDMS slab (outlined in lavender) after detachment is complete. The layer of cells, after peeling, is a smaller flat rectangle outlined in maroon dots. Scale bar 1000$\mu$m. (C) Schematic of the cell sheet peeled from PDMS with parallel stripes (top) and plain (bottom). For the striped pattern $x_i$ and $y_i$ are the initial dimensions of the PDMS sample parallel and perpendicular to the stripes, respectively. $x_f$ and $y_f$ are the final dimensions of the peeled sheet along the corresponding axes. For the plain PDMS $x_i$ and $y_i$ are defined as the dimensions of the longer and shorter axes of the PDMS, respectively, and $x_f$ and $y_f$ are the final dimensions of the peeled cell sheet along the corresponding axes. (D) Contraction of the cell sheet parallel and perpendicular to the alignment of the cells. Sample data from plain PDMS (here, $C_\parallel$ indicates the contraction along the long axis of the sample) are indicated in orange squares and and data for striped PDMS are indicated in purple triangles. The black dotted line corresponds to isotropic contraction ($C_\parallel=C_\perp$).  Errors are described in Methods. The white square and triangle indicate the contraction for isotropic or aligned samples, respectively, obtained via simulation. (E) Final shape of simulated cell sheets with stripes (top) and without (bottom) and their height profiles. The color indicates the vertical deviation of the sheets with both remaining largely flat. }
    \label{fig1}
\end{figure}

We grow NIH-3T3 mouse fibroblasts to a high density on slabs of polydimethylsiloxane (PDMS), treated with poly-D-lysine, a common polypeptide for cell culture. At very high density, after a few days of growth, a flexible skin made of fibroblasts and extracellular matrix tends to spontaneously detach from the substrate~\cite{Bose2019, Bischofs2008}, peeling off starting from the edges of the sample. Any perturbations to the environment, such as shaking or changing growth media, triggers such detachment. A similar effect can also be induced by degradation of the substrate gel~\cite{Elvitigala2023} or with the use of thermoresponsive polymers~\cite{okano_pnipam}. In our experiments, we drag a scalpel blade along the perimeter of the slab. As soon as the edges of the cell layer are released from the PDMS, the detachment of the whole layer proceeds spontaneously throughout the sample (Fig.~\ref{fig1}A and Supplemental Fig. S1). Upon detachment, we notice that most fibroblasts are in the floating sheet and only very few fibroblasts are left on the PDMS substrate. The area of the detached cell sheet is significantly smaller than the initial area occupied by the cells on the substrate (Fig.~\ref{fig1}B), showing auxetic behavior with contraction along both dimensions in 2D.

Fig.~\ref{fig1}A shows a phase contrast image of cells (microscope Nikon Ti-2 Eclipse) during peeling from PDMS substrates patterned with parallel ridges, or stripes, 10$\mu$m wide and 1.5-2$\mu$m tall, spaced by 60$\mu$m. We observe that often the sheet starts to detach from one edge and to lift over the course of several minutes (Supplemental Fig. S1). The peeling speed varies between samples and is often stalled by points where cells are pinned to the surface. An example of this is shown in Supplemental Fig. S2. The dynamic appears to be highly variable: sometimes the detachment starts from one edge or corner, and the pinning points can alter the course of the peeling front.

Despite the variability in the dynamics, the final peeled sheet demonstrates an elastic characteristic, consistently having a final shape matching that of the substrate from which it detached. In Fig.~\ref{fig1}B, shown as a typical example, the detached cell sheet (outlined in maroon dotted line) has a different size and aspect ratio from the substrate (outlined in lavender) but it shares the features of being a rectangular shape, albeit with jagged edges, with the corners of the detached sheet and the substrate maintaining substantial alignment.

This consistency allows us to compare the degree of contraction obtained from plain substrates and from striped substrates. On the stripes NIH-3T3 cells reach confluency with uniform nematic alignment, with the director oriented along the stripes (also shown in Supplemental Fig. S1.A). We can see from microscopy that although there are some height fluctuations across the sample, the cell sheets remain essentially flat. Sometimes the samples' corners may fold inward creating a deformation in the margin, due to the movement of the liquid above the floppy cell sheet. The sheets that fully roll into tubes or tear due to the fluid movement are discarded in our analysis.

If the sheet detaches without rolling, as in Fig.~\ref{fig1}B, we can measure the contraction from 2D phase contrast images, combined together with ImageJ stitching plugin~\cite{Preibisch2009}. We take as reference the dimensions of the PDMS to determine the initial dimensions of the cell sheet ($x_i$ and $y_i$) and measure the final dimensions of the sheet ($x_f$ and $y_f$), as shown in schematics of Fig.~\ref{fig1}C. Without loss of generality, we choose the $\hat{x}$ direction to be aligned with the stripes.

We estimate the fractional contraction along the nematic director ($C_\parallel$) and the fractional contraction perpendicular to the director ($C_\perp$) as $C_\parallel=\frac{x_i-x_f}{x_i}$ and $C_\perp=\frac{y_i-y_f}{y_i}$, respectively. The results are shown in Fig.~\ref{fig1}D (purple triangles), where each point represents the measured total contraction for one cell sheet and the error bars indicate uncertainties in the measurements that arise from their irregular shapes (as detailed in Methods). The data consistently show that the cell sheets that peel from striped PDMS undergo a greater contraction along the direction parallel to the stripes. The average $C_\parallel$ observed among these samples is $0.7 \pm 0.04$, whereas the average $C_\perp$ is $0.49 \pm 0.07$. The difference is apparent from the visible change in the aspect ratio of the samples.

We can compare this with cells grown on plain rectangular blocks of PDMS, which do not impose any anisotropy onto the cells. Here we define $C_\parallel$ as the contraction along the longest edge of the sample. In the absence of stripes, we observe isotropic contraction ($C_\parallel\approx C_\perp$), see~\ref{fig1}D (orange squares). The average $C_\parallel$ among these samples is $0.66 \pm 0.05$, whereas the average $C_\perp$ is $0.69\pm 0.07$. From this we conclude that the cell sheets are isotropically contractile, with an additional anisotropic contraction caused by the alignment. The anisotropic contraction suggests the cell sheet behaves as an anisotropic elastic material. We then define the material's stretch factor as $\lambda = 1-C_\parallel$, which gives the factor by which the material extends/contracts in the direction parallel to the cellular alignment. The stretch factor perpendicular to the alignment can be written as $1- C_\perp = \lambda^{-\nu}$ where we have introduced the Poisson ratio, $\nu$. 


The data points in Fig.~\ref{fig1}D are taken by varying parameters such as the initial density of cells (the seeding density), the growth time and the initial area of the cell sheet, the aspect ratio of the PDMS slab on which the cell grow, as detailed in Supplemental Table 1. Based on our data, none of these parameters show a clear influence on $C_\parallel$ or $C_\perp$, at least within our errors. However, we cannot completely exclude, for example, that the initial aspect ratio of the slab may influence the contraction (this was suggested in~\cite{Serbo2016}). Finally, we can also test the viability of cells after detachment. As shown in Supplemental Fig. S3, if the detached cell sheet is deposited upon a petri dish, the cells will start to grow, showing that the cells remain alive during the process.


We can simulate the contraction of the cell sheets by considering them as thin elastic materials in which the local relaxation is anisotropic, depending on the orientation of the alignment, see SI and~\cite{pearce2020defect} for details. We simulate this elastic material as a network of springs, with a rest length that expands or contracts depending on the local orientation of the cells. We introduce microscopic local contractions which are fitted to the data in Fig.~\ref{fig1}D, see methods section for details. 
Fig.~\ref{fig1}E shows the initial and final shape of a simulated cell sheet both with (top) and without (bottom) anisotropy. The black rectangular contour indicates the initial size and the purple lines (where present) indicate the orientation of the ridges. The surfaces are colored by their vertical height with the $XZ$ and $YZ$ cross sections alongside, indicating no curvature is generated with or without stripes. Following the same protocol as in experiments, we measure $C_\parallel$ and $C_\perp$ in simulations, giving estimates of $C_\parallel \approx 0.69$ and $C_\perp \approx 0.51$ shown in Fig.~\ref{fig1}D. This corresponds to values of $\lambda = 0.31$ and $\nu =-0.609$ for our simulated cell sheets.

\begin{figure}[hb!]
    \centering
    \includegraphics[width=0.80\textwidth]{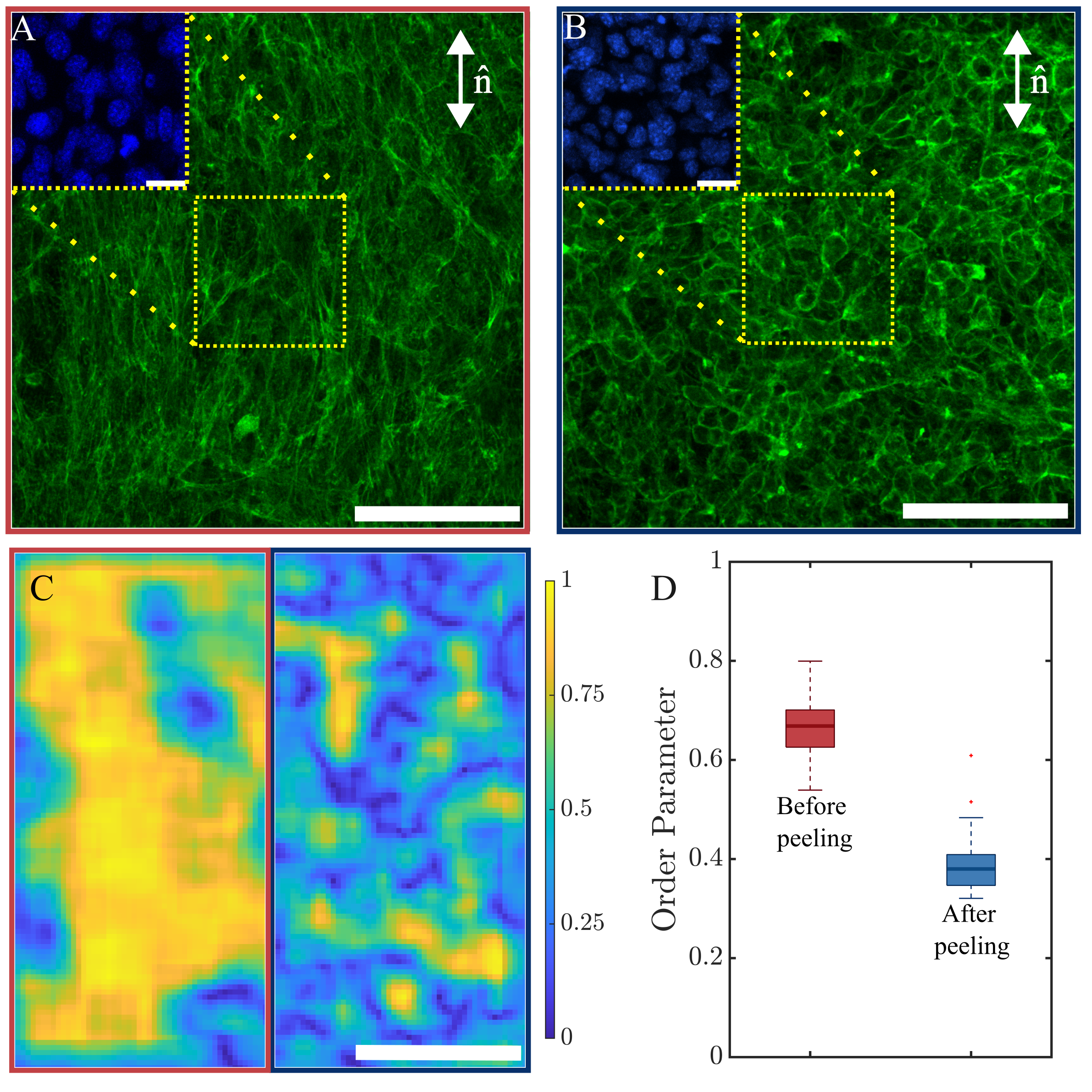}
    \caption{\textbf{Order parameter change upon detachment.} (A-B) Fluorescence image of actin filaments and nuclei (stained with Phallodin for actic and NucBlue$^\mathrm{TM}$ Fixed Cell ReadyProbes$^\mathrm{TM}$ for nuclei, in insets) before (A) and after (B) the peeling of cells. Scale bars 100$\mu m$ in figure and 20$\mu m$ in insets. (C) Color maps of order parameter $S$ measured from images (A) on the left and (B) on the right after the analysis with OrientationJ (only half the images are shown). (C) Box plot of order parameter as defined in the paper, and calculated over 2 different samples (total n=21 images) before peeling and 2 different samples (total n=31 images) after detachment.}
    \label{fig2}
\end{figure}

\subsection*{Nematic order}


In liquid crystal elastomers, $\nu$ is the opto-thermal Poisson ratio and it relates the parallel and perpendicular changes in length during phase transitions. We would like to verify that also for the cells the anisotropic contraction corresponds to a change in the degree of nematic order. We quantify the degree of order of the cell layer on the striped substrate before and after detachment by looking at the alignment of actin filaments. We image the actin filaments stained with phalloidin rhodamine, and nuclei stained with NucBlue$^\mathrm{TM}$ Fixed Cell ReadyProbes$^\mathrm{TM}$ and analyze the fluorescence microscopy images with OrientationJ~\cite{Püspöki2016}. The results are shown in Fig.~\ref{fig2}, where we see the fluorescence images of the actin network before (Fig.~\ref{fig2}A) and after (Fig.~\ref{fig2}B) detaching, alongside the nuclear fluorescence (insets). The alignment direction of the ridges ($\hat{n}$) is shown in the figure, and it is evident that the actin appears more uniformly aligned in panel A, with the alignment direction matching that of the stripes. However, this alignment is lost when the cells are peeled from the substrate, and there is no clear correlation between the direction of the director and actin filaments. Nuclei also show a more tightly packed cell layer, as the observed nuclear density is higher in detached cell sheets compared to the cell layers still attached on the substrates. 

We quantify the degree of alignment of the actin with a color map (Fig.\ref{fig2}C) representing an order parameter $S$ defined as : $$ S= \sqrt{\langle \cos 2 \theta \rangle_{x,y\in Q} ^2 + \langle \sin 2 \theta \rangle_{x,y\in Q} ^2} $$ where $\theta$ is the angle between vectors detected by OrientationJ within a square region $Q$ (following~\cite{Duclos2016}). The $S$ of cell layers on the striped substrates reduces after detachment as shown in Fig~\ref{fig2}D. 

When the cells are supported by the textured substrate, it induces a degree of nematic alignment in the cells, indicated by the increased $S$. When the cells are released from the substrate, the cells relax to a less aligned state, indicated by the drop in $S$. To approach an isotropic state, the cells contract more along their long axis, leading to anisotropic contraction. Thus we consider the peeling process itself as a way to induce a transition between a higher order and a lower order phase, or to simplify between nematic and isotropic phase.



\subsection*{Actuation of cell layers}\label{part3}
\begin{figure}[h!]
    \centering
    \includegraphics[width=0.80\textwidth]{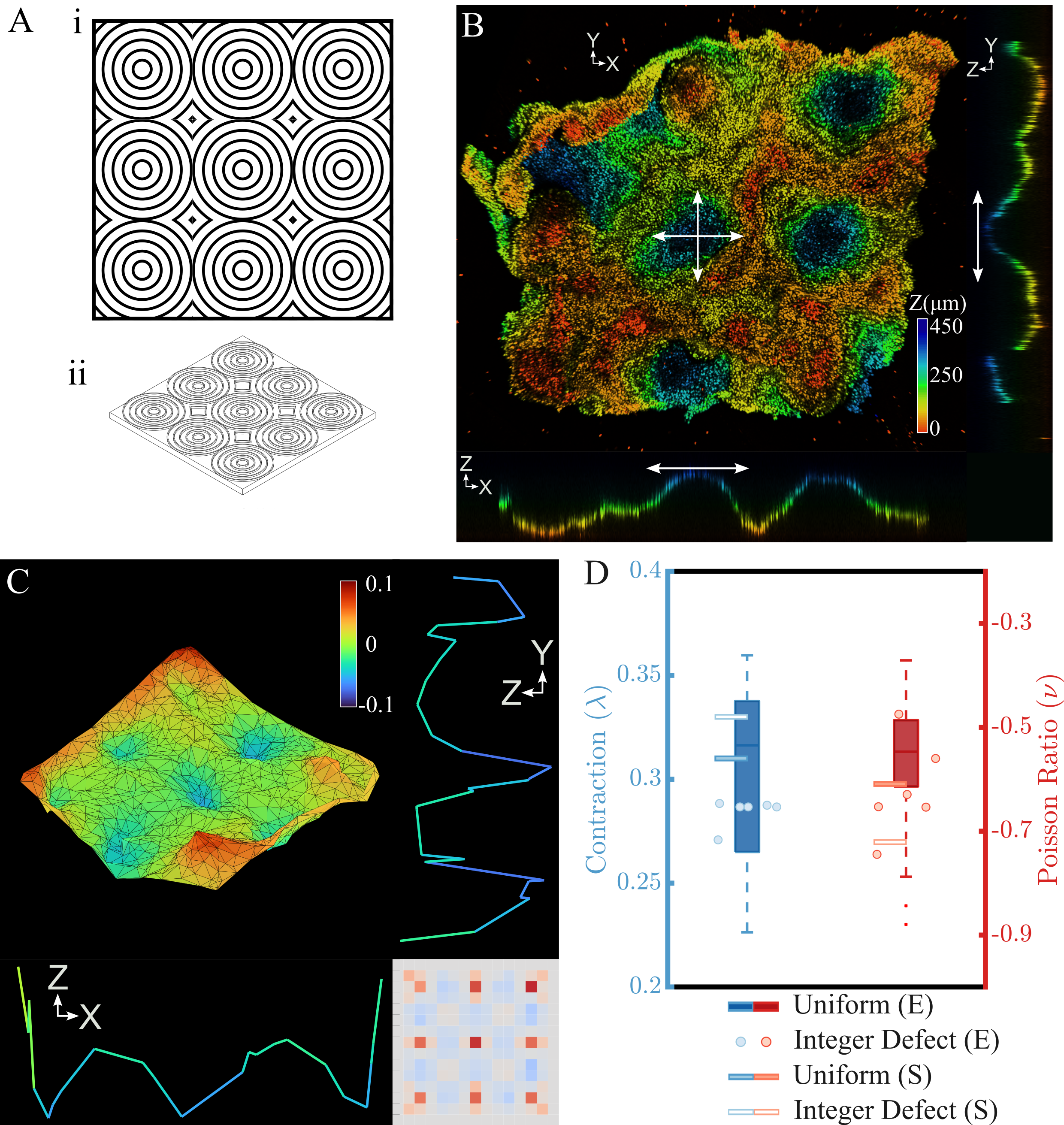}
    \caption{\textbf{Actuation of cell sheets.} Schematic of the defect array patterns used in experiments (i), and a 3D rendering of the substrate topography with the height of the ridges at 1.5-2$\mu$m, ridge width 9$\mu$m (ii). (B) Confocal microscopy image of a sheet of cells detached from the defect array. The overall dimensions of the peeled sample are 2000$\mu$m x 2500$\mu$m. The color bar shows the height difference. XZ and YZ cross-sections along the directions indicated by white arrows reveal a sheet of uniform thickness with localized out-of-plane deformations. (C) Corresponding simulations of the cell sheet deformation. The inset on the bottom right corner represents a color map of Gaussian curvature (see SI for details). (D) Values of stretch factor $\lambda$ and Poisson ratio $\nu$ calculated from experiments (box plots for the flat samples with parallel ridges, and clear circles from curved samples with +1 topological defects) and simulations (filled rectangles for flat samples and clear rectangles for samples with +1 topological defects).}
    \label{fig3}
\end{figure}

Since the cell sheets from striped substrates exhibit anisotropic contraction, as expected with a change of order parameter, we also test whether spatially varying alignment can generate Gaussian curvature. We first study a 2D array of +1/-1 topological defects, arranged in a square lattice, represented with an approximate schematic in Fig.\ref{fig3}A, which we have extensively studied in previous work~\cite{Endresen2021, Kaiyrbekov2023, ravichandran2025topology}. PDMS is cut into 3$\times$3 grids of +1 defects (total size 7.5mm x 7.5mm), which introduces four -1 defects in between. It is known that in liquid crystal elastomers this pattern generates cones when actuated upon phase transition~\cite{Ware2015}.

Fig.~\ref{fig3}B is a confocal microscopy image showing the structure of a cell sheet detached from a +1/-1 defect array, in which the cell nuclei were tagged with NucBlue$^\mathrm{TM}$ Live Cell ReadyProbes$^\mathrm{TM}$ (Hoescht 33342) stain. The peeled sheet shows $3\times3$ regions of localized positive Gaussian curvature, bearing a striking resemblance to the cones formed by actuated liquid crystal elastomers. The cross sections confirms that curved cell sheet remains thin, even in the mounds, thus they are not formed by accumulation of cells (Fig.~\ref{fig3} and S4). This result is highly reproducible and other samples are shown in SI. 

Using the parameters we obtained from the flat sheet, we can also simulate a cell sheet featuring a $3\times3$ defect array. As in experiments, we observe mounds located at the center of each topological defect, see Fig.~\ref{fig3}C. The $XZ$ and $YZ$ projections show slices through the central topological defect, and we see that both principal curvatures have the same sign indicating positive Gaussian curvature. We can directly calculate the polyhedral Gaussian curvature of the simulated surface, which shows a clear $3\times3$ grid of localized positive Gaussian curvature, see Fig.~\ref{fig3}C (bottom right inset).

The generation of Guassian curvature around a +1 topological defect can simply be understood by considering the change in radius and circumference of the defect. Since the circumference is parallel to the alignment of the cells, we expect it to contract more than the radius. This change in the ratio between the circumference and radius of the patch of cells induces positive Gaussian curvature. For an infinitely thin material with constant $\lambda$ and $\nu$, the predicted final shape is a cone. The steepness of the cone can be related to the contraction of the material by the equation $\sin(\alpha) = \lambda^{1+\nu}$ where $\alpha$ is the half angle of the cone, see~\cite{Modes2011} for a derivation.

Instead of cones, however, both the cell sheets and simulations feature mounds with smoother curvature, as can be seen from the cross-section in Fig.~\ref{fig3}B and Supplemental Fig.~S4. This is not surprising and is due to three factors. First, the cell layer has a non-negligible thickness, implying a finite bending energy which causes the cone tip to be smoother (as shown for example in~\cite{Guin2018}). Furthermore, the cell layer is soft, making it difficult to sustain the discontinuous shape of a cone. Moreover, the size of the defect core in cell monolayers is large compared to the small (ideally null) size of the topological defect core in liquid crystal elastomers. Finally, also in simulations the defect core has zero $S$ by definition, which implies anisotropic relaxation is reduced toward the center of a defect, thus $\lambda$ and $\nu$ are not spatially constant.

\begin{figure}[h!]
    \centering
    \includegraphics[width=0.80\textwidth]{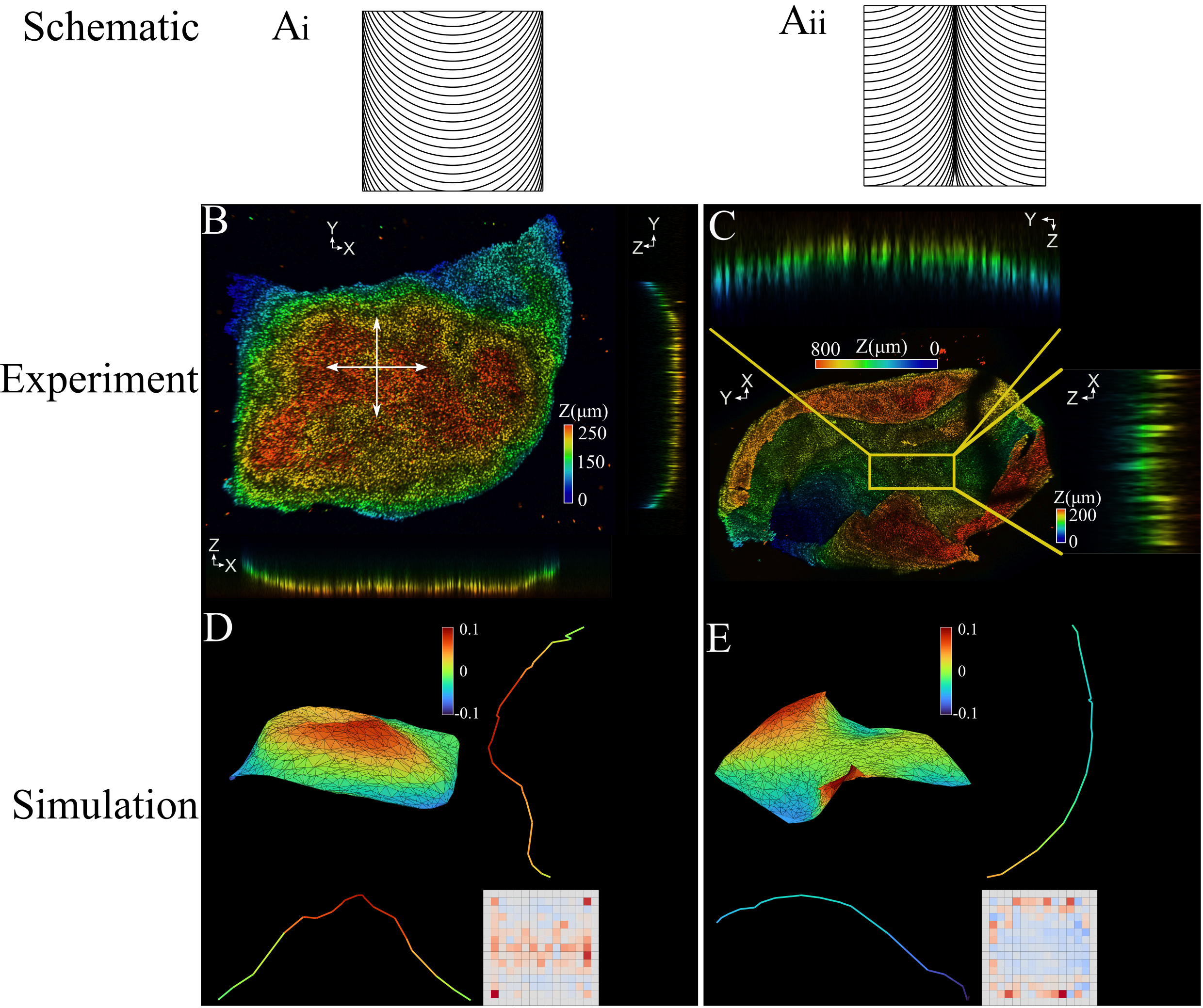}
    \caption{\textbf{Actuation of cell sheets with non-singular patterns.} (A) Schematics of the tested alignment patterns featuring  bend (left), and splay (right) patterns. All the patterns vary in 2-D and are made of short ridges like the schematic in \ref{fig3}A (ii) with the height of the ridges at 1.5-2$\mu$m, ridge width 9$\mu$m. (B-C) Confocal microscopy image of a peeled sheet of cells from the bend (B), and splay (C) patterns. The overall size of the detached samples are 1800$\mu$m x 3000$\mu$m (B), and 1800$\mu$m x 3000$\mu$m (C). The color bar shows the height difference. The cross sections show the consistent thickness of the sheet, and the curvature along the two principal directions for the bend (B) and splay (C) patterns, respectively. (D-E) Same as previous row but for simulated cell sheets, akin to what was seen in \ref{fig3}C. The insets in the corner are the maps of the Gaussian curvature of the sheets averaged over multiple simulations, see SI for details.
    } 
    \label{fig4}
\end{figure}

We can estimate an effective value of $\lambda$ and $\nu$ from the confocal microscopy images and compare them to those measured for the flat sheets. Fig.~\ref{fig3}D shows the box plots of $\lambda$ and $\nu$ for cell sheets with uniform alignment, while the filled horizontal rectangles show the result of the simulations of a flat sheet.
For a mound generated by a topological defect, we can estimate $\lambda$ by taking the ratio between the circumference of the largest circular ridge around a topological defect in the initial pattern, and the circumference of a fully formed mound. The latter is calculated in experiment from the distance between two adjacent mound tops multiplied by $\pi$. This gives estimates of $\lambda$ highly compatible with those that we measured for flat sheets. We can estimate the half angle of the cone, $\alpha$, by directly measuring the steepness of the mounds, see Fig.~\ref{fig3}B, which we then combine with our estimate of $\lambda$ to give an estimate of the Poisson ratio, which is again in good agreement with our measurements from flat sheets (purple circles Fig.~\ref{fig3}D). We take the same protocol with our simulated cell sheets and obtain estimates of $\lambda = 0.33$ and $\nu = -0.72$, again highly compatible with our estimates from experiments and simulated flat sheets. Note that $\nu$ is negative, as was also reported for other biological tissues such as tendons, arteries, and skin~\cite{Timmins2010, Gatt2015, Mardling2020}, reflecting that the actuation occurs by a contraction in both x and y. This auxetic material is unlike conventional liquid crystal elastomers, which typically expand in the directions perpendicular to the director.



The in-plane contraction of the material dictates the presence of positive Gaussian curvature, but does not imply a particular sign of mean curvature. This means that a localized peak or valley are equally likely to form at the center of the +1 topological defect. In liquid crystal elastomers, the bending direction is typically determined by the position of the heating source or a tilted angle of molecular alignment on the surface~\cite{Aharoni2018}. In our experiments and simulations we observe bending in both directions within the same sheet, as shown in Supplemental Fig. S5. This implies that symmetry is not broken between the two sides of the cell sheet. 

While all experimental and simulation results shown so far are consistent with a liquid crystal elastomer-like description, the higher density of cells at +1 topological defects~\cite{Kawaguchi2017, Endresen2021, Kaiyrbekov2023, Zhao2024} may also play a role in generating the curvature~\cite{Pfeifer2024}. Since the cells in the core of the +1 defects are packed to a higher density and already have a smaller size compared to cells in the surrounding sheet, it is possible that they cannot contract as much as the cells further away from the defect core. Furthermore, it has been shown that relaxing the liquid crystal energy around a +1 topological defect can also lead to cone shapes~\cite{pearce2025passive}. To rule out these alternative mechanisms and demonstrate that the cells' 3D structure is a consequence of their nematic order, we test two non-singular patterns with constant cell density and order parameter that are able to generate Gaussian curvature in liquid crystal elastomers. These patterns are shown in Fig.~\ref{fig4}A (i-ii) and were previously studied by Mostajeran and coworkers~\cite{Mostajeran2016}. We verify that we can indeed generate positive or negative Gaussian curvature in the center of the samples shown in Fig.~\ref{fig4} B, and C respectively, and depicted from simulations in Fig.~\ref{fig4} D, and E, as expected. In experiments, the edges of the samples tend to roll randomly, most likely due to the very soft and floppy nature of the elastic material. Nevertheless, this measurement shows that the actuation occurs and generates Gaussian curvature, even without the variation of cell density and order parameter associated with topological defects.

\section*{Discussion}

We have demonstrated that during detachment, aligned sheets made of NIH-3T3 fibroblasts contract anisotropically. Their contraction is greater along their nematic director than perpendicular to it, with a stretch factor $\lambda\approx0.3$ and an opto-thermal Poisson ratio of $\nu \approx -0.55$. Because of this anisotropic contraction, we can generate 3D structure by spatially patterning the alignment of the cells in 2D. This is demonstrated in cell sheets patterned both with +1/-1 defect arrays and with non-singular patterns (Fig.~\ref{fig3}). All these results are supported by numerical simulations. Finally, the detached cell sheet is viable immediately after peeling and can be transferred to a new dish for regrowth. 

These results show that, as for liquid crystal elastomers, it is the in-plane alignment that determines the structure of the peeled cell sheet. Curved cell sheets can be used as precursors for spheroids and more complex 3D assemblies. Curvature control is usually achieved by deforming the substrate of the cell culture~\cite{Kim2024}. Here we propose a route to control curvature through the control of the 2D alignment. Our simulations give excellent agreement with experiments, and give a rapid way to relate 2D alignment patterns to final 3D structures. More studies are needed to clarify the role of cell density and extracellular matrix development, but our work identifies a promising direction for fundamental studies of tissues as metamaterials that can achieve target shapes thanks to the nematic order, and sets a step in linking nematic order to tissue morphogenesis.


\section*{Methods}\label{Methods}
\subsection*{Cell culture} 
NIH 3T3 mouse fibroblast cells (ATCC) are used in all experiments. The cells are cultured on Thermo Fisher Scientific Nunclon$^{TM}$ Delta surface-coated dishes in 89\% Dulbecco’s Modified Eagle’s Medium (DMEM) - high glucose formulation (containing 4.5 g/L glucose, L-glutamine, sodium pyruvate, and sodium bicarbonate; Sigma Aldrich), supplemented with 10\% fetal bovine serum (FBS; Sigma Aldrich) and 1\% penicillin-streptomycin. The cells used in the experiments are between passages G9 and G15.

\subsection*{Substrate manufacturing}
Substrates are manufactured in a clean room using standard photolithography techniques on 10 cm silicon wafer chips. SU-8 2 negative photoresist is used to reach the desired height of  1.5 $\mu$m to 2 $\mu$m for the designs and measured using a profilometer.

\subsection*{Substrate preparation}
Polydimethylsiloxane (PDMS; Sylgard 184, Dow Corning), mixed with 10\% curing agent, is used as a substrate for cell experiments. An SU-8-coated silicon wafer is placed in a petri dish and coated with a thin layer (3–5 mm) of desiccated PDMS mixture (1:10 ratio). The petri dish is then desiccated again to remove any bubbles formed during pouring. After desiccation, the sample is cured on a hot plate at 90$^{\circ}$C for 3–4 hours. Once cured, the PDMS substrate is carefully removed.

This PDMS mold is then used to create negative molds with Norland Optical Adhesive 81 (NOA-81), a UV-curable glue. A flat-bottomed glass petri dish containing 3–4 drops of NOA-81 is prepared, and the PDMS mold was inverted onto it to ensure complete coverage of the patterned substrate with NOA-81. The sample is degassed to ensure accurate replication of the pattern by the UV glue. The petri dish is exposed to UV light at 302 nm (8 W; Ultra Violet Products-3UV) for 20 minutes. The petri dish is then flipped, and an additional 20 minutes of UV exposure is done to ensure complete polymerization of the UV glue. Finally, the dish is heated at 60$^{\circ}$C for 30 minutes. Once completed, the original PDMS mold is removed, leaving behind a negatively patterned substrate made of UV glue, which is then used to produce PDMS copies for experiments.

The patterned PDMS copies (parallel ridges, or integer defects) are cut into required sizes and cleaned using scotch tape. A thin layer (2 mm) of PDMS (1:10) is poured onto a plastic petri dish and cured overnight at 37$^{\circ}$C. The petri dish along with the pattered PDMS substrate is treated with oxygen plasma (Harrick Plasma Cleaner) with RF power 30W for 3 minutes with a pressure of 300mtorr. Once the plasma-cleaned, the PDMS slab is attached to the plastic petri dish with the patterned side facing up and then heated for 1 min at 60$^{\circ}$C. 

The sample is then sterilized with ethanol and prepared for substrate treatment with Poly-D-Lysine hydrobromide (Sigma-Aldrich). The substrate is coated with 0.1 mg/ml concentration of Poly-D-Lysine in Milli-Q Water, barely covering the surface. After 30 mins the dish is cleaned with sterile Milli-Q Water and allowed to dry for 90 mins. Once dried the cells are planted on the substrate. The cell concentration of the suspension is determined using a hemocytometer with 10 $\mu$l demarcation. The concentration of cells used for plating the dish varies between 500-1000 cells/mm$^2$. 

\subsection*{Peeling method}
Cells are allowed to grow undisturbed for 2-4 days inside an incubator at 37$^{\circ}$C with 5\% CO$_2$. Once the cells are over-confluent, the samples are carefully carried to the microscope so as not to induce spontaneous peeling. Once at the microscope, we use a scalpel with the sharp side of the blade touching the edge of the slab but pointing opposite to the scratching direction. With this, we scratch each edge of the PDMS slab. This is carried out for all edges, from one vertex to another ensuring the bonds of the cells from the edge of PDMS are carefully severed, while also not cutting into the PDMS. Once cut, the cells either spontaneously peel completely from the surface or get pinned at times as shown in Fig. S2. In this case, a sharp-end tweezer is used to cut the anchored cell, while observing through the microscope. To aid in peeling the same tweezers are also used to brush the cell sheet which is already peeling, while ensuring not to puncture the cell sheet. This results in a completely detached cell sheet from the substrate. The cells are slightly denser than water, so the cell sheet tends to settle on top of the substrate. For staining the cell sheet for actin, the cell sheets are pinned close to a corner of the cell sheet through to the substrate while peeling with a 50 $\mu$m needle, while being observed under the microscope. This helps prevent any movement of cell sheet while the staining protocol is carried out.

\subsection*{Fixing and staining}
To observe the cell nuclei, NucBlue$^\mathrm{TM}$ Live Cell ReadyProbes$^\mathrm{TM}$ (Hoescht 33342) stain is added to the dish at a concentration of 1 drop/ml of cell media and followed by 30 mins of incubation. 

Cells are also fixed and observed before and after peeling. For post-peeling samples, as described in Peeling Method, a 50 $\mu$m needle is used to secure the unattached cell sheet to the substrate. All samples undergo the same fixation protocol, using 4\% Paraformaldehyde (PFA) in PBS Ready-to-Use Fixative (Biotium) to fix the cells sheet. Around 100 $\mu$l of PFA is applied over a  4 mm$^2$ cell sheet and incubated for 20 mins. The sample is then washed twice with 1x Phosphate-buffered saline (PBS), leaving the PBS in the sample for 5 mins during each wash. 

For actin staining, cells are permablilized by adding 50-100$\mu$l of 0.1\% (V/V) Triton X-100 (Merck) in 1x PBS for 10 mins, follwed by two 5-minute washes with 1x PBS. To block non-specific binding 50-100$\mu$l of 1\%(W/V) Bovine serum albumin (BSA, VWR) in 1x PBS is added and incubated for 1 hour, after which the samples are washed twice more with PBS 1x. Actin is tagged using  0.2\% (V/V) Rhodamine Phallodin (invitrogen) diluted in 1x PBS. 50-100 $\mu l$ of this solution is used to coat the same and incubated for 1 hour before washing it with again with 1x PBS. Following this, the cell sheet is stained using NucBlue$^\mathrm{TM}$ Fixed Cell ReadyProbes$^\mathrm{TM}$, adding 2 drop/ml to 1x PBS in the petri dish, incubated for 20 mins and washed again with 1x PBS. 

\subsection*{Microscopy}
Nikon Tl-Eclipse Widefield microscope and Kinetix Scientific CMOS camera (Teledyne Photometrics) are used to image cell sample, using phase contrast and fluorescent imaging in 2D. Large multi-point image is used to capture the complete cell sheet and patterned substrate by translating the stage along a grid with 15\% overlap between the frames.

Confocal imaging is done using the Nikon A1R confocal unit with Ti-2 LFOV micorscope body, and with A1-DUG hybrid 4-channel multi detector. LU-NV laser unit is used to excite the sample at 407nm and 514nm for nuclei and actin respectively.

\subsection*{Distance and error estimation}
Images obtained from both the microscope are processed by using ImageJ Stitching plugin by passing them through an ImageJ Stitching plugin~\cite{Preibisch2009}. For Z-Stack large image stitching, an ImageJ macro was written and used in conjuction with ImageJ Stitching Plugin to get final image. ImageJ was later used to the measure the various length of the cell sheet and the ridges.

To account for the irregular shape of the cell sheets before and after peeling, we take 5 measurements per side of the samples before and after peeling, evenly distributed across the sample. In some cases, samples have a partially rolled side, or an ``tear'' near a corner. These are clearly visible from the phase contrast images. In this case, we measure the local thickness of the rolled side and add it to the measured length.  

The percentage contraction in x and y directions are calculated using
\begin{align}
    C_X=\frac{x_i-x_f}{x_i} \nonumber \\
    C_Y=\frac{y_i-y_f}{y_i}\label{SI_contraction_equation}
\end{align}
where $x_i, y_i, x_f \,\mathrm{and} \, y_f$ are the average initial (substrate or before peeling dimensions) and final (post peeling dimensions) of the cell sheet. Errors in estimation denoted by $\delta x_i, \delta y_i, \delta x_f \,\mathrm{and} \, \delta y_f$ is given by standard deviation of measured $x_i, y_i, x_f \,\mathrm{and} \, y_f$.
With this the error in $C_X \, \mathrm{and} \, C_Y$ is given by 

\begin{align}
    \delta C_X=\pm \biggl\{ \frac{\delta x_f +\delta x_i}{x_i - x_f}+\frac{\delta x_i}{x_i} \biggl\}\frac{x_i-x_f}{x_i} \nonumber \\
    \delta C_Y=\pm \biggl\{ \frac{\delta y_f +\delta y_i}{y_i - y_f}+\frac{\delta y_i}{y_i} \biggl\}\frac{y_i-x_f}{y_i} 
    \label{SI_uncertainity_equation}
\end{align}

\subsection*{Order parameter}
In order to get a projection of a confocal image, ImageJ is used to get a Maximum intensity projection of actin images. The maximum intensity image is passed through OrientationJ (ImageJ plugin) with a local window size ($\sigma$ = $5 \mu m$) to get the orientation of actin fibers, both before and after the peeling. The orientation is then used to calculate the local order parameter $S$~\cite{Duclos2016}

\begin{align}
    S=\sqrt{\langle \cos 2 \theta \rangle_{x,y\in Q} ^2 + \langle \sin 2 \theta \rangle_{x,y\in Q} ^2} \label{SI_Orderparamter_Equation}
\end{align}

where $\theta$ is the local nematic director. $S$ varies from 0 to 1 and captures the degree of nematic alignment within the designated area. The local order parameter is taken over a $30 \mu$m window.



\bibliographystyle{ieeetr}





\bigskip
\bigskip



\newpage

\setcounter{figure}{0}
\renewcommand{\figurename}{Fig.}
\renewcommand{\thefigure}{S\arabic{figure}}

\section*{Supplementary Information}\label{sec5}

\subsection*{Supplementary Figure}\label{SI_Fig}

\begin{figure}[ht!]
    \centering
    \includegraphics[width=1\textwidth]{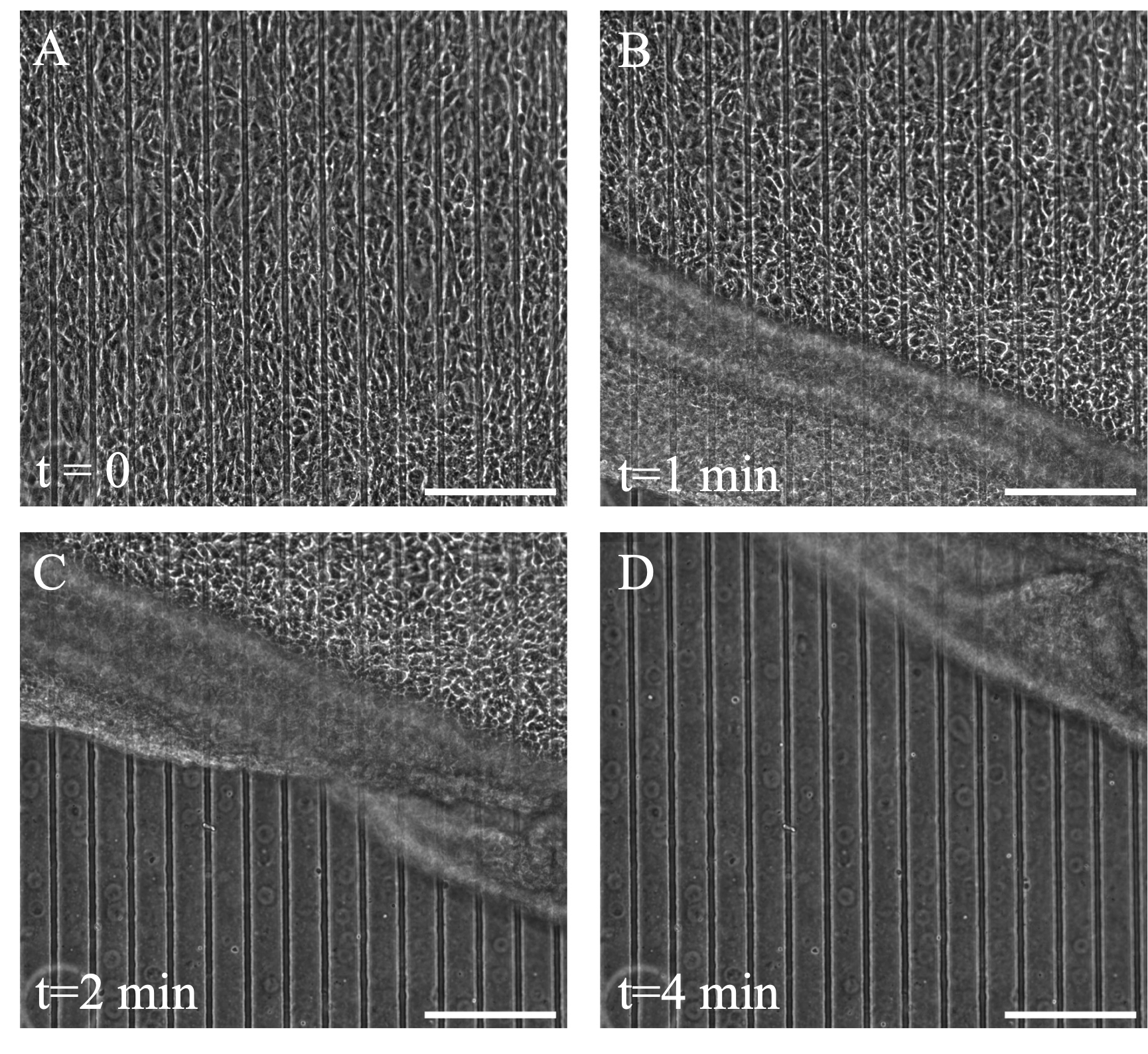}
    \caption{\textbf{Progression of peeling cell sheet over 4 minute period.} (A) At \textit{t=0}, or just before scraping the edges of the cell layer, cells are attached to the PDMS substrate, which is patterned with h=2$\mu$m stripe ridges with w=60$\mu$m spacing. The cells are aligned and elongated along the direction of the stripes. (B) At \textit{t=1}min after scraping the edges, the cells are starting to peel from the substrate. The peeling front has reached the middle of the image. The fully lifted part of the cell sheet is visible near the bottom of the image (C) At \textit{t=2}min, the peeling front has traveled to the top of the image. (D) After \textit{t=4}min, the cells within the frame are almost fully peeled. Scale bars are 200$\mu$m.}
    \label{fig:peelingtimeseries}
\end{figure}


\begin{figure}[h!]
    \centering
    \includegraphics[width=0.8\textwidth]{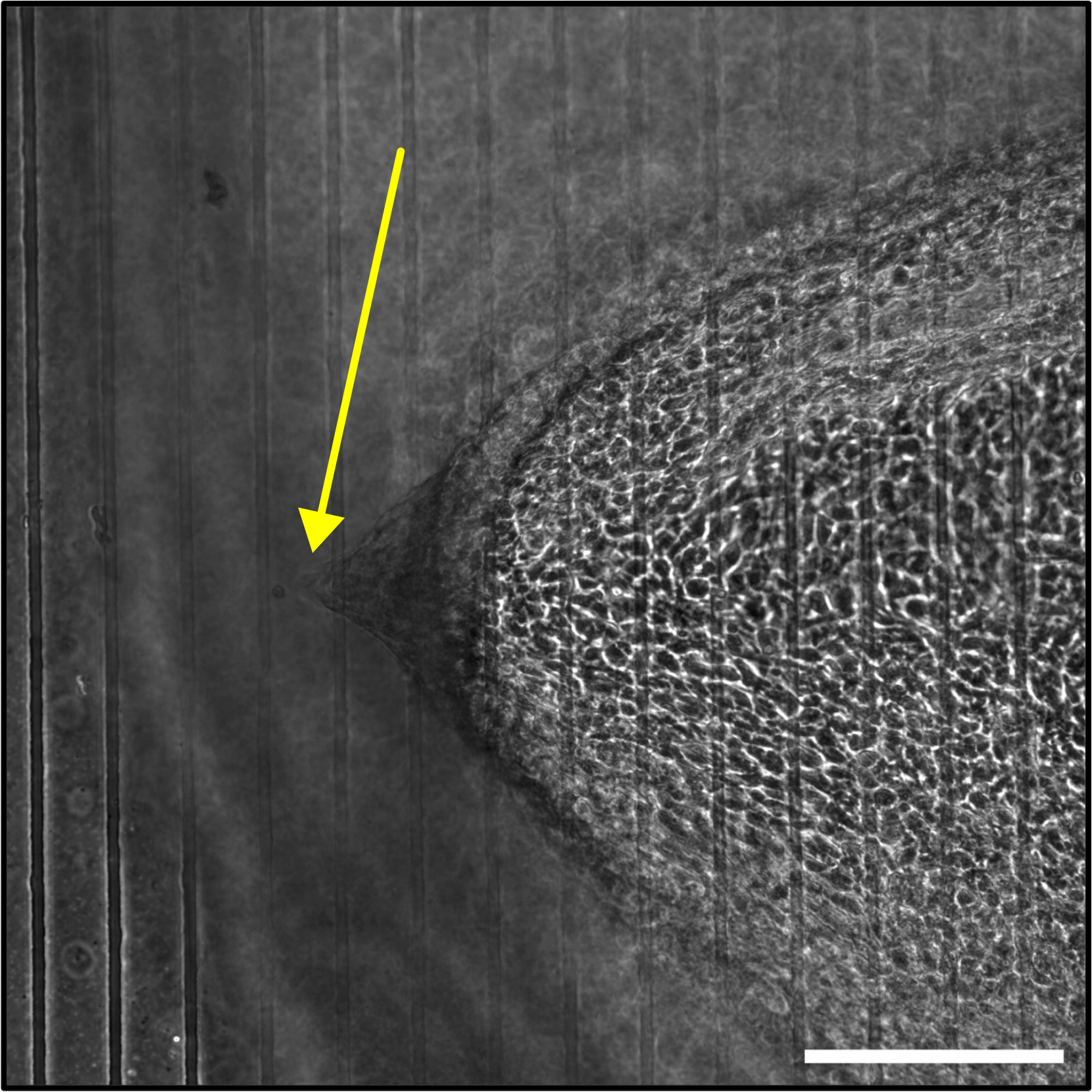}
    \caption{\textbf{Pinning points, locations where the cell sheet is strongly pinned to the PDMS, occur during the peeling process.} Here, the peeling front is traveling from the left of the image. The cell sheet remains attached to the substrate at this pinning point, as the peeling front deforms around it. The pinning point is observed through the lifted part of the cell sheet and is indicated by the arrow. The part of the cell sheet to the right of the pinning point has not yet peeled. Scale bar is 200$\mu$m.}
    \label{fig:pinningPoint}
\end{figure}

\begin{figure}[h!]
    \centering
    \includegraphics[width=0.8\textwidth]{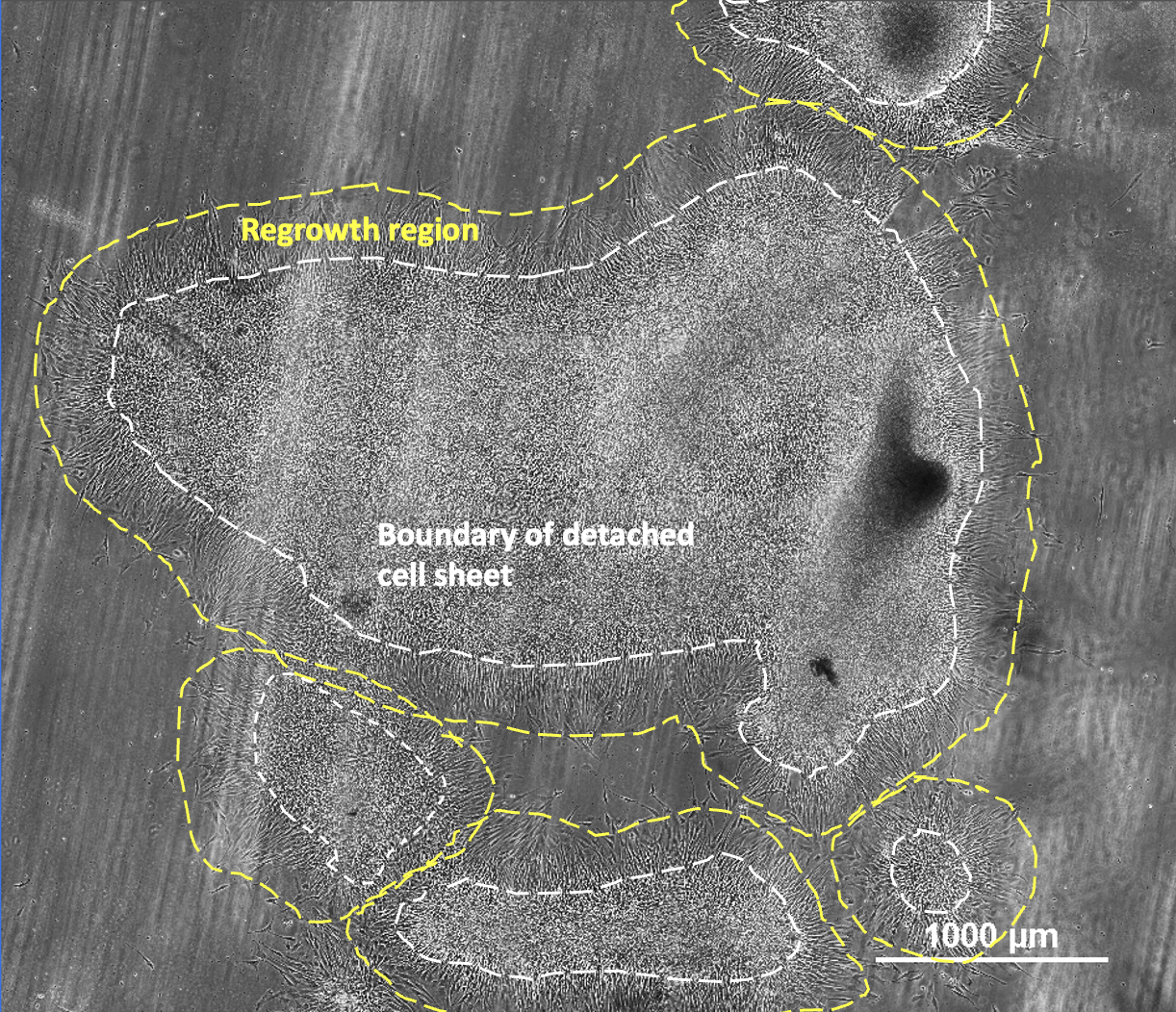}
    \caption{\textbf{Regrowth of cells after detachment.} A previously detached cell sheet reattaches and grows in a plastic Petri dish coated with fibronectin. Cells in the regrowth region are oriented radially outward from the cell sheet. The image is obtained after 1 day of regrowth inside a cell incubator microscope stage (Okolab Stage Top Incubator H-301K). The boundary of the detached cell sheets are identified as regions with high density of compact cells. The figure demonstrates that the cells are viable after peeling.}
    \label{fig:regrowth}
\end{figure}

\begin{figure}[h!]
    \centering
    \includegraphics[width=0.8\textwidth]{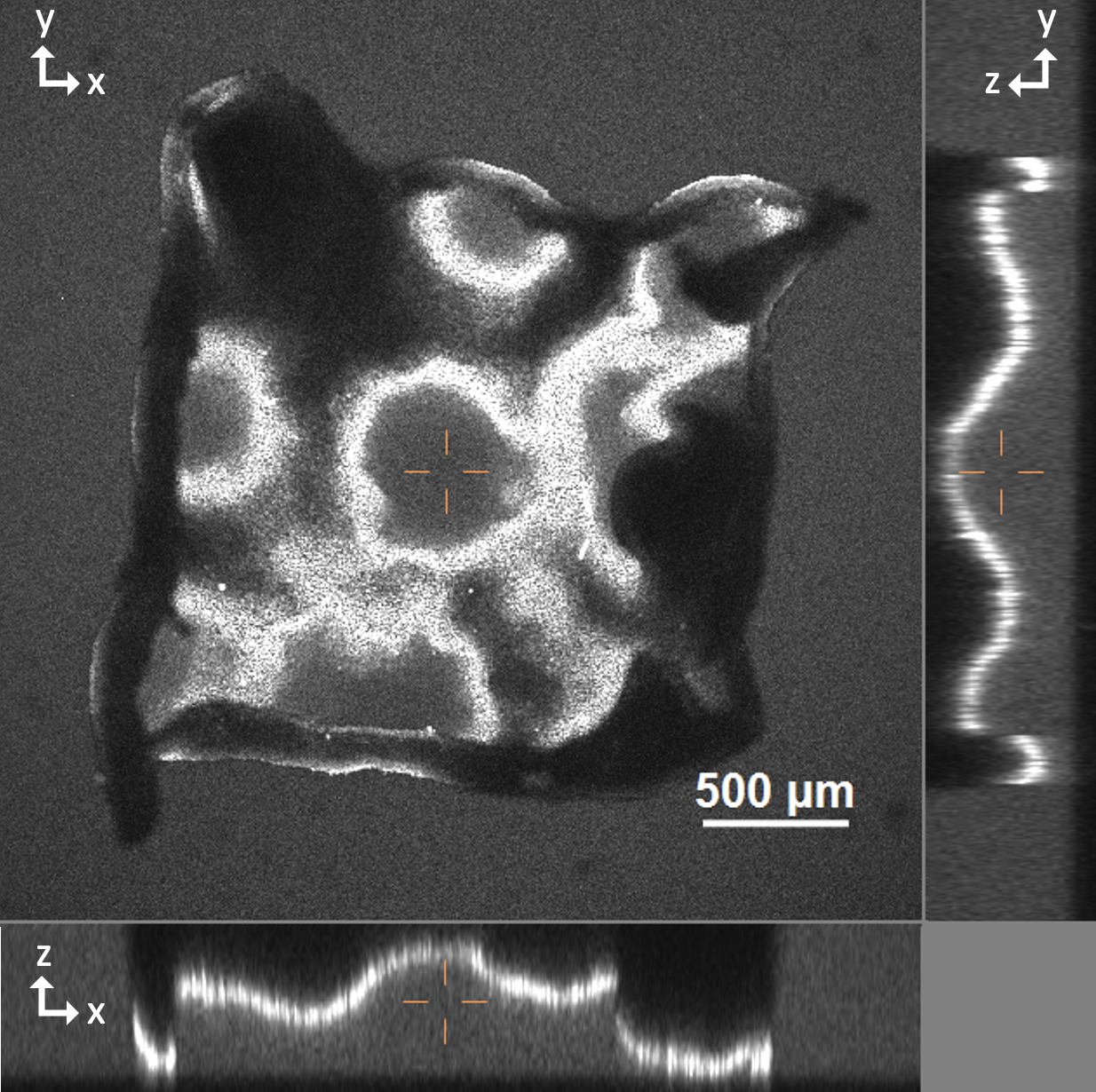}
    \caption{Confocal microscopy image of a cell sheet after peeling off from a defect array pattern, shown in Fig. 3A.i of the main paper. On the sides, the lateral projections are shown, measured along the lines indicated by the cross in the figure. From the projection it is possible to see that the cell sheet, evident as the bright white profile remains a thin layer after peeling.}
    \label{fig:confocal_ortho}
\end{figure}

\begin{figure}[h!]
    \centering
    \includegraphics[width=0.8\textwidth]{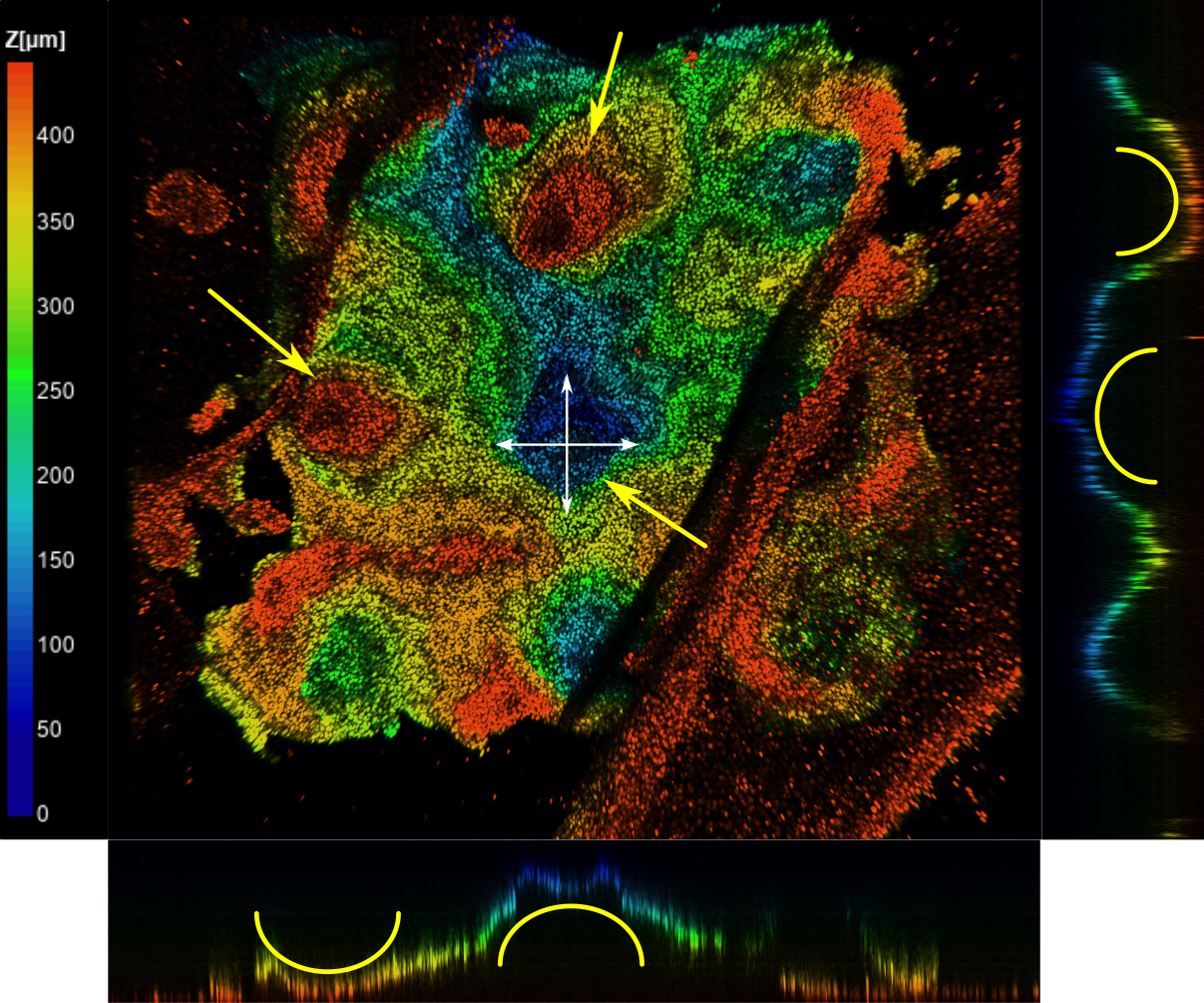}
    \caption{Confocal microscopy image of a cell sheet after peeling off from a defect array pattern, shown in Fig. 3A.i of the main paper. The 3D projection of the image illustrates the cusp and bowl at the location of +1 topological defect, indicated by the yellow arrows. On the sides, the lateral projections are shown, measured along the lines indicated by the cross in the figure. The two projections clearly show a cusp and a bowl (indicated by yellow half circles) suggesting it is random with the out-of-plane deformation of the cell sheet.}
    \label{fig:cusp_bowl}
\end{figure}

\clearpage
\subsection*{Supplementary Table}\label{SI_Table}


\begin{table}[h]
\caption{Parameters for the data points (orange squares) in Fig. 1D for cell layers peeled from plain PDMS slabs. We report: the seeding density of the cells in cells/mm$^2$, the days of growth before peeling, the aspect ratio of the PDMS slab, and the contractions parallel and perpendicular to the x direction.}\label{tab1}
\begin{tabular*}{\textwidth}{@{\extracolsep\fill}lcccccc}
\toprule%
Sample & Seeding Density & Days of Growth & Aspect Ratio & $C_\parallel$ & $C_\perp$ \\
\midrule
\midrule
Sample 1  & 500 & 3 & 1.2 & $0.60\pm0.08$ & $0.60\pm0.05$ \\
Sample 2  & 1000 & 3 & 1.5 & $0.62\pm0.02$ & $0.66\pm0.03$ \\
Sample 3  & 500 & 4 & 1.5 & $0.66\pm0.04$ & $0.70\pm0.05$ \\
Sample 4  & 1000 & 3 & 1.5 & $0.70\pm0.03$ & $0.75\pm0.03$ \\
Sample 5  & 1000 & 3 & 1.5 & $0.71\pm0.04$ & $0.77\pm0.04$ \\
\bottomrule
\end{tabular*}
\end{table}

\begin{table}[h]
\caption{Parameters for the data points (purple triangles) in Fig. 1D for cell layers peeled from striped PDMS slabs. We report: the seeding density of the cells in cells/mm$^2$, the days of growth before peeling, the aspect ratio of the PDMS slab, and the contractions parallel and perpendicular to the nematic director.}\label{tab2}
\begin{tabular*}{\textwidth}{@{\extracolsep\fill}lcccccc}
\toprule%
Sample & Seeding Density & Days of Growth & Aspect Ratio & $C_\parallel$ & $C_\perp$ \\
\midrule
\midrule
Sample 1  & 1000 & 3 & 1.5 & $0.76\pm0.02$ & $0.57\pm0.05$ \\
Sample 2  & 1000 & 2 & 1.5 & $0.77\pm0.01$ & $0.52\pm0.05$ \\
Sample 3  & 500 & 3 & 1.5 & $0.74\pm0.02$ & $0.53\pm0.07$ \\
Sample 4  & 500 & 4 & 1.5 & $0.74\pm0.01$ & $0.56\pm0.05$ \\
Sample 5  & 500 & 3 & 2 & $0.64\pm0.03$ & $0.58\pm0.09$ \\
Sample 6  & 500 & 3 & 1 & $0.70\pm0.04$ & $0.60\pm0.04$ \\
Sample 7  & 500 & 3 & 1.6 & $0.68\pm0.03$ & $0.59\pm0.04$ \\
Sample 8  & 500 & 3 & 1.5 & $0.65\pm0.03$ & $0.61\pm0.03$ \\
Sample 9  & 500 & 3 & 2.8 & $0.70\pm0.02$ & $0.48\pm0.07$ \\
Sample 10  & 500 & 3 & 2.5 & $0.73\pm0.03$ & $0.38\pm0.08$ \\
Sample 11  & 500 & 3 & 2.5 & $0.75\pm0.03$ & $0.49\pm0.07$ \\
Sample 12  & 500 & 3 & 2.9 & $0.68\pm0.02$ & $0.36\pm0.07$ \\
Sample 13  & 500 & 3 & 2.8 & $0.69\pm0.01$ & $0.46\pm0.10$ \\
Sample 14  &500 & 4 & 2 & $0.67\pm0.02$ & $0.44\pm0.05$ \\
Sample 15  &500 & 4 & 1.9 & $0.66\pm0.03$ & $0.41\pm0.03$ \\
Sample 16  & 500 & 4 & 1.4 & $0.66\pm0.02$ & $0.48\pm0.04$ \\
Sample 17  & 500 & 4 & 1.6 & $0.66\pm0.02$ & $0.48\pm0.04$ \\
Sample 18  & 500 & 4 & 1.5 & $0.66\pm0.03$ & $0.45\pm0.05$ \\
Sample 19  & 500 & 4 & 1.6 & $0.66\pm0.02$ & $0.48\pm0.04$ \\

\bottomrule
\end{tabular*}
\end{table}

\subsection*{Simulation Details}

The simulation approximates the cell sheet with a point cloud which are joined by springs that approximate the internal stresses of the material. Over the course of the simulation, the rest lengths of the springs are adjusted to reflect the changing internal stresses. The point cloud is allowed to relax this stress which results in the changing shape of the cell sheet. 

The motion of the points in the point cloud follow the overdamped Langevin equation:
\begin{equation}
    \dot{\underline{x}}_i = \mu \underline{F}_i
    \label{eq:v}
\end{equation}
where $\underline{F}_i$ is the force on point $i$.

The force on point $i$ is calculated as the sum of all spring forces acting on $i$, which is written:
\begin{equation}
    {\underline{F}}_i = k\sum_{j\in C_i}\underline{\hat{r}}_{ij}(l_{ij} - l^*_{ij})
    \label{eq:F}
\end{equation}
where $\underline{\hat{r}}_{ij}$ is the unit vector pointing from point $i$ to point $j$, $k$ is the spring constant, and $l_{ij}$ is the distance from point $i$ to point $j$. 

$C_i$ is the set of points connected to point $i$ by a spring, which is determined by performing a 3D Delaunay triangulation on the point cloud. Finally, $l^*_{ij}$ is the rest length of the spring connecting point $i$ and $j$. The rest length is calculated according to the following equation. 

Next, a Delaunay triangulation is used to identify which points are connected by springs. We calculate the final rest length of each spring according to the following equation:
\begin{equation}
    l^* = l^0\sqrt{\alpha(1 + (\beta-\gamma/2) \Delta S + \gamma \Delta S(\hat{\underline{l}}^0\cdot\hat{\underline{P}})^2)}
    \label{eq:l}
\end{equation}
Where $l^0$ represents the initial length of the spring, $\underline{\hat{l}}^0$ is the unit vector parallel to the initial spring. $\underline{\hat{P}}$ is a unit vector parallel to the nematic texture at the midpoint of the spring in its initial position, $\Delta S$ is the change in the order parameter at the midpoint of the spring.

The coefficients $\alpha$, $\beta$ and $\gamma$ control how the rest length of the spring changes over the course of the simulation. $\alpha$ corresponds to a isotropic, uniform rescaling of the system, $\beta$ corresponds to an isotropic rescaling of the system that depends on the local order parameter, and $\gamma$ corresponds to an anisotropic rescaling of the system parallel to the nematic texture. Note here, $\gamma$ is defined to be area preserving rescaling which is the reason that it occurs twice. 

The stripes are described by the fields $\underline{\hat{P}}$ and $\Delta S$, which we consider to depend on the $x$ and $y$ positions within the initial sheet only.  

To initialize the simulation, $1000$ points are placed at random within a thin, flat sheet with dimensions $L_x$, $L_y$ and $L_z$. The positions of the points are then adjusted to reduce the coulomb energy associated with the point cloud, this is to ensure the points are roughly evenly spread out in three dimensions. The Delaunay triangulation is then performed on this arrangement of points and $l^*$ is calculated for every spring according to Eq.~\ref{eq:l}.

The forces on the points are calculated using Eq.~\ref{eq:F} which are used to iterate the positions of the points according to Eq.~\ref{eq:v}. We rescale our time units by $k\mu$ and integrate Eq.~\ref{eq:v} using a Euler scheme with a timestep $\Delta t = 5\times 10^-2$. For simulations used in Fig.~\ref{fig1} we set $L_x = 2L_y = 100L_z = 1$ and in Fig.~\ref{fig3} and ~\ref{fig4} we set $L_x = L_y = 100L_z = 1$. Simulations were run for $10^4$ simulation steps in all cases. Maps of Gaussian curvature provided in Fig.~\ref{fig3} and ~\ref{fig4} were calculated by averaging over 10 simulations with different initial conditions. 

\subsubsection*{Simulation Parameter Fitting}

We first take the uniform contraction in the absence of stripes to be $C_\parallel = 0.66\pm 0.05$ and $C_\perp = 0.69\pm 0.07$ from Fig.~1 in the manuscript. We simulate this configuration by setting $\Delta S=0$, which leaves $\alpha$ as the only remaining free parameter. 
We can predict the value of $\alpha$ by fitting $C_\parallel$ with Eq.~\ref{eq:l} the following equation:
\begin{equation}
    C_\parallel = 1 - l/l_0 = 1-\sqrt{\alpha}
\end{equation}
Where a value of $\alpha=0.1$ corresponds to a prediction of $C_\parallel = 0.68$.

In simulation with $\alpha=0.1$ and $\Delta S = 0$, we obtain $C_\parallel = 0.68636  \pm 0.00069$ and $C_\perp= 0.68803  \pm 0.00122$ in good agreement with experiments, see Fig.~\ref{fig1}C.

By setting $\underline{\hat{P}} = [1,0]$ we introduce the uniform stripe pattern from Fig.~1 in the manuscript; in experiments, this corresponds to a change of $C_\parallel\approx0.7$ and $C_\perp \approx 0.50$. We fit each of these two data points individually. However, now we have three fitting parameters since $\Delta S\neq0$, this means that the solution is not necessarily unique.
\begin{align}
    C_\parallel &= 1-\sqrt{\alpha(1 + \Delta S\beta + \Delta S\gamma/2)} \approx 0.7\\
    C_\perp &= 1-\sqrt{\alpha(1 + \Delta S\beta - \Delta S\gamma/2)} \approx 0.5
\end{align}
We retain the general rescaling parameter $\alpha = 0.1$, which, along with the first equation gives $\Delta S\beta = -\Delta S\gamma/2$. This can be interpreted as fitting the 2D Poisson ratio of the change in shape. The second equation now allows us to fit $\Delta S\gamma = -1.5$. Finally, we take $\Delta S = 0.3$ from Fig.~\ref{fig2}D which can be used to fix $\gamma$ and $\beta$.

Simulating with these parameters gives good agreement with the experimental data, see Fig.~\ref{fig1}D. More precisely, we get $C_\parallel=0.68608  \pm 0.00091$ and $C_\perp=0.50590 \pm 0.00072$ 

\subsubsection*{Nematic Patterns for Simulation}

The nematic patterns used in the simulations are as follows. 

Fig.~\ref{fig1}F. $\underline{P} = [1,0]$

Fig.~\ref{fig3}, ~\ref{fig4}. We define a director angle $\theta$ such that $\underline{P} = [\cos(\theta),\sin(\theta)]$. We take $\Delta S = 0.3$ everywhere. 

For the defect array (Fig.~\ref{fig3}), we divide the area into a $3\times3$ grid, the center of each square contains a defect. The defect array texture is the given by $\theta = \phi + \frac{\pi}{2}$ where $\phi$ is the polar angle around the center of the current grid square. The local change in order parameter is given by $\Delta S = 0.3 - \sum_i\exp(-r_i/\epsilon)$ where the sum runs over the location of the center of each defect and $\epsilon = L_x/100$ is the defect core radius. 

For the bend texture (Fig.~\ref{fig4} left) the texture is simply given by $\theta = -\pi/2 + \frac{x\pi}{L_x}$. We take $\Delta S = 0.3$ everywhere. 

For the splay texture (Fig.~\ref{fig4} right) the texture is given by $\theta = \frac{x\pi}{L_x}$. We take $\Delta S = 0.3$ everywhere. 

\subsubsection*{Measurement of Contraction and Poisson Ratio in simulation}
We measured the contraction $\lambda$ and Poisson ratio $\nu$ for the defect array in Fig.~\ref{fig3} in a similar manner as done for the experiments, here for the center cone of each simulation. $\lambda$ is measured as the contraction of the whole surface which is the same as the contraction of the circle around the center cone. $\nu$ is calculated from estimating the mound as cone as in Ref.~\cite{Modes2011} with the half opening angle $\alpha$ with $\sin{\alpha} = \lambda^{1+\nu}$. We set a target radius at which we measure the height of the cone. This is done for different azimuthal angles around the cone. Only if the discretization point distance and the chosen target distance closely match, the $\nu$ is taken into account.
For different simulations, values can be found in Tab.\ref{tab:sim_defect_array_lambda_nu}. On average we found $\lambda=0.328$, $\nu=-0.718$.

\begin{table}[h]
\caption{Contraction $\lambda$ and Poisson ratio $\nu$ for different simulations.}
\centering
\begin{tabular*}{0.3\textwidth}{@{\extracolsep\fill}cc}
   $\lambda$ & $\nu$ \\
    0.328 & -0.699 \\
    0.329 & -0.69 \\
    0.346 & -0.933\\
    0.316 & -0.646\\
    0.322 & -0.622  
\end{tabular*}    
\label{tab:sim_defect_array_lambda_nu}
\end{table}

\end{document}